\documentclass[flushrt]{emulateapj}

\begin{document}

\title{Two Lensed Lyman-$\alpha$ Emitting Galaxies at \lowercase{z}$\sim$5}

\author{
Matthew B. Bayliss\altaffilmark{1,2}, 
Eva Wuyts\altaffilmark{1,2}, 
Keren Sharon\altaffilmark{2}, 
Michael D. Gladders\altaffilmark{1,2}, 
Joseph F. Hennawi\altaffilmark{3}, 
Benjamin P. Koester\altaffilmark{1,2}, 
H{\aa}kon Dahle\altaffilmark{4}
}

\email{mbayliss@oddjob.uchicago.edu}

\altaffiltext{*}{Based on observations obtained at the Gemini
Observatory, which is operated by the Association of Universities
for Research in Astronomy, Inc., under a cooperative agreement
with the NSF on behalf of the Gemini partnership: the National
Science Foundation (United States), the Science and Technology
Facilities Council (United Kingdom), the National Research Council
(Canada), CONICYT (Chile), the Australian Research Council (Australia),
Minist\'{e}rio da Ci\^{e}ncia e Tecnologia (Brazil) and Ministerio de
Ciencia, Tecnolog\'{i}a e Innovaci\'{o}n Productiva (Argentina), and
the Apache Point Observatory 3.5-meter telescope, which is owned and
operated by the Astrophysical Research Consortium.}
\altaffiltext{1}{Department of Astronomy \& Astrophysics, University of Chicago, 5640 South Ellis Avenue, Chicago, IL 60637}
\altaffiltext{2}{Kavli Institute for Cosmological Physics, University of Chicago, 5640 South Ellis Avenue, Chicago, IL 60637}
\altaffiltext{3}{Max-Planck-Institut f{\"u}r Astronomie K{\"o}nigstuhl 17, D-69117, Heidelberg, Germany}
\altaffiltext{4}{Institute of Theoretical Astrophysics, University of Oslo, P.O. Box 1029, Blindern, N-0315 Oslo, Norway}

\begin{abstract}

We present observations of two strongly lensed $z\sim5$ Lyman-$\alpha$
Emitting (LAE) galaxies that were discovered in the Sloan Giant Arcs Survey
(SGAS). We identify the two sources as SGAS J091541+382655, at $z=5.200$,
and SGAS J134331+415455 at $z=4.994$. We measure their AB magnitudes at
$(i,z)=(23.34\pm0.09,23.29\pm0.13$) mags and $(i,z)=(23.78\pm0.18,24.24^{+0.18}_{-0.16}$)
mags, and the rest-frame equivalent widths of the Lyman-$\alpha$ emission at
$25.3\pm4.1$\AA~and $135.6\pm20.3$\AA~for SGAS J091541+382655 and
SGAS J134331+415455, respectively. Each source is strongly lensed
by a massive galaxy cluster
in the foreground, and the magnifications due to gravitational lensing are recovered
from strong lens modeling of the foreground lensing potentials. We use the
magnification to calculate the intrinsic, unlensed Lyman-$\alpha$ and UV continuum
luminosities for both sources, as well as the implied star formation rates (SFR).
We find SGAS J091541+382655 and SGAS J134341+415455 to be galaxies with
(L$_{Ly-\alpha}$, L$_{UV})\leq(0.6$L$_{Ly-\alpha}^{*}, 2$L$_{UV}^{*}$) and
(L$_{Ly-\alpha}$, L$_{UV})=(0.5$L$_{Ly-\alpha}^{*}, 0.9$L$_{UV}^{*}$),
respectively. Comparison of the spectral energy distributions (SEDs) of both sources
against stellar population models produces estimates of the mass in young stars in each
galaxy: we report an upper limit of M$_{stars} \leq 7.9^{+3.7}_{-2.5} \times 
10^{7}$ M$_{\sun} h_{0.7}^{-1}$ for SGAS J091531+382655, and a range of
viable masses for SGAS J134331+415455 of $2\times10^{8}$ M$_{\sun} h_{0.7}^{-1} <$
M$_{stars} < 6\times10^{9}$ M$_{\sun} h_{0.7}^{-1}$.

\end{abstract}


\section{Introduction}

Understanding the evolution of galaxies -- especially the first generation of
galaxies -- remains one of the most important topics in astrophysics and cosmology.
Populations of galaxies in the distant universe are identified over a wide
range of wavelengths, including Distant Red Galaxies (DRGs), Ultra Luminous
Infra-Red Galaxies (ULIRGs) and Sub-Millimieter Galaxies (SMGs).
Many efforts to study the properties of high redshift galaxies at optical wavelengths
focus on two distinct classes selected by their rest-frame UV properties:
1) Lyman-Break Galaxies (LBGs) and 2) Lyman-$\alpha$
Emitters (LAEs).  LBGs are selected via deep wide-band photometry, identified
by the `Lyman limit' continuum break that appears at 912\AA~ in
the rest frame \citep{steid1996a,steid1996b,lowenthal1997} -- though for sources
at higher redshift this spectral break moves redward, approaching 1216\AA~ in
the rest frame due to the Lyman-$\alpha$ forest \citep{steid1987,rauch1998}
absorption by intergalactic neutral hydrogen. LAEs are selected by either
narrow-band imaging \citep{cowie1998,rhoads2000,rhoads2003,ajiki2004,gawiser2006,yamada2005}
or blind spectroscopy \citep{kurk2004,sawicki2008} tuned to detect Lyman-$\alpha$
line emission redshifted into near-ultraviolet, optical, or near-infrared wavelengths.
Over the past decade large samples of
LBGs and LAEs have driven studies of star-forming galaxies at $z\gtrsim2.5$.

Surveys for LBGs and LAEs are efficient for collecting statistical samples of
high-redshift galaxies, but at $z\gtrsim3$ they produce objects which are generally
too faint to have their galactic continuum emission studied spectroscopically. The
standard approach for studying
the properties of these galaxy samples relies on stacking the photometric signal
from many objects and fitting the observed mean SED against a variety of stellar
population synthesis models in order to constrain parameters such as the ages and masses
of the underlying stellar populations, as well as the amount of dust extinction
\citep{shapley2003,chary2005,pirzkal2007,lai2007,fink2008,fink2009a,nilsson2009,yabe2009}.
In principal, stellar population synthesis modeling can also provide information
about dust properties (i.e. the shape of the dust law) and metallicity, but even the
stacked SED signal at $z\gtrsim3$ is  insufficient to constrain these additional
parameters with much confidence. Broadly speaking, galaxies selected as LBGs are believed
to sample more massive star-forming galaxies with an underlying older stellar
population, and possibly higher dust content, while LAE selected galaxies tend to be
lower mass galaxies with low metallicities and very little dust
\citep{giaval2002,venem2005,gawiser2007}.
Hubble Space Telescope imaging studies of high redshift LAE galaxies imply that
these sources are compact, and likely either disk-like or irregular in structure
\citep{pirzkal2007,tanigu2009}. Recently \citet{fink2009c} modeled individual
SEDs of 14 bright $z\sim4.5$ LAEs from the Chandra Deep Field South and found a broad
range in stellar population age, stellar mass, and dust extinction, which suggests
that stacking SED analyses of high redshift galaxies may not be the best approach.

The main hurdle involved in studying any high redshift source is the general
lack of signal. Distant galaxies are faint and therefore difficult to
detect, and those which are identified are rarely -- if ever -- amenable to
detailed follow-up. Furthermore, those sources which are bright enough to be
studied individually are drawn from the extreme
bright tail of the luminosity function of high redshift galaxies, and are
therefore not necessarily representative of the bulk of the populations.
In this paper we present two serendipitously discovered, strongly lensed high redshift
galaxies: \object{SGAS J091541+382655}, spectroscopically confirmed at $z=5.200\pm0.001$,
with $r_{AB}=24.68\pm0.25$, $i_{AB}=22.92\pm0.09$ and
$z_{AB}=22.75\pm0.13$ mags, and \object{SGAS J134330+415455} spectroscopically confirmed at
$z=4.994\pm0.001$, with $r_{AB}\geq25.47$ mags, $i_{AB}=23.36\pm0.18$ mags
and $z_{AB}=23.70^{+0.18}_{-0.16}$ mags. Both objects have $riz$ colors and magnitudes
that satisfy selection criteria for $r-$band dropouts in $z\sim5$ dropout surveys,
as well as
Lyman-$\alpha$ equvialent widths (see Section 2) sufficiently large to be
selected in surveys for Lyman-$\alpha$ excess. At $z\gtrsim5$ these objects are
the two brightest LAEs in the literature to date, and both sources are projected
on the sky within ~$30\arcsec$ of the cores of
confirmed strong lensing galaxy clusters. This means that the sources -- once
corrected for the lensing magnification -- are intrinsically much fainter than
the observed flux implies, and therefore provide a rare opportunity to study
individual LAE properties at the fainter end of the luminosity function. There is a
small but growing number of magnified galaxies at high redshift that are excellent
candidates for high resolution spectroscopic follow-up \citep{koester2010,wuyts2010};
some of these galaxies have been observed in detail at optical and
near-infrared wavelengths,
including cB58 \citep{pettini2000}, "the 8 O'clock Arc" \citep{allam2007,fink2009b},
and "the Cosmic Eye" \citep{smail2007,siana2009,quider2010}. The two galaxies
discussed in this paper are the first $z\sim5$ galaxies that present similar
opportunities for detailed study via follow-up spectroscopy.

Where necessary we calculate cosmological distances assuming a flat cosmology with
$H_{0}=70$ km s$^{-1}$ Mpc$^{-1}$, and matter density $\Omega_{M}=0.3$. All magnitudes
are AB.

\section{Observations}

\subsection{Data}

The two sources presented here were first identified as $\emph{r-}$band dropouts
in $\emph{gri}$ imaging of two different strong lensing galaxy clusters,
and subsequently confirmed by spectroscopy to have strong Lyman-$\alpha$
emission features. The two galaxy clusters were identified as part of the
Sloan Giant Arcs Survey \citep[SGAS;][]{hennawi2008}, a blind survey for strong
lensing systems in optically selected massive clusters at $0.1\leq z \leq 0.6$
detected via the Red-Sequence Cluster algorithm \citep{gladders2000} adapted to
run on the Sloan Digital Sky Survey \citep[SDSS;][]{york2000} public data
release catalogs. Strong lensing clusters are identified by visual
inspection of imaging in $g-$band on 2m to 4m-class telescopes for
600s in $<1\arcsec$ seeing, and the most spectacular systems have been
followed-up with multi-band imaging and spectroscopy on 8m-class telescopes.
One of the clusters discussed here, SDSS J1343+4155, appears in a recent small
sample of strong lensing clusters discovered in the SDSS by \citet{diehl2009}.
The other cluster, SDSS J0915+3826, does not appear in any prior published work.
Imaging and spectroscopic observations were conducted with the Frederick C. Gillett
Telescope (Gemini North) between the months of 2008 February and 2008 July.  The
GMOS imaging observations were pre-imaging conducted for the purpose of spectroscopic
mask design. Both imaging and spectroscopy were part of Gemini program GN-2008A-Q-25.
The primary goal of the spectroscopic observations was to obtain redshifts
of arcs to facilitate strong lensing modeling. We briefly summarize the spectroscopy
here, and refer the reader to Bayliss et al. (2010), in preparation, for additional
details.

\begin{figure}
\centering
\includegraphics[scale=0.225]{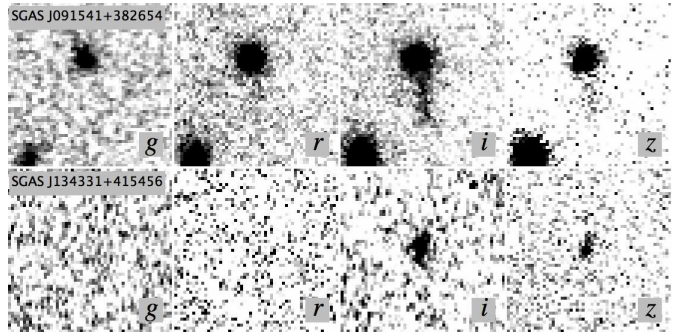}
\caption{\scriptsize{GMOS $\emph{griz}$ (from left to right)
$8\arcsec\times8\arcsec$ cutout images centered on SGAS J091541+386254
and SGAS J134331+415455. Both objects exhibit obvious drop-out behavior
from the $i-$ to $r-$bands. The $z-$band detection of SGAS J091541+386254
appears at lower significance than its counterpart for SGAS J134331+415455
because its $z$ data were taken during bright time and
the sky background noise is $\sim30\%$ higher.}}
\label{dropout}
\end{figure}

Gemini North/GMOS $\emph{gri}$ photometry for both cluster fields are derived
from 2x150s dithered exposures, which were executed in queue mode on February and
March of 2008. These pre-imaging data were used for spectroscopic mask design and
target prioritization. GMOS $\emph{z-}$band observations consisted of 6x180s
dithered exposures which were scheduled as follow-up, primarily in order to
measure the continuum flux of the LAEs, and were executed in queue mode in February
of 2010. The GMOS images were reduced using the Gemini IRAF\footnote{IRAF (Image
Reduction and Analysis Facility) is distributed by
the National Optical Astronomy Observatories, which are operated by AURA, Inc.,
under cooperative agreement with the National Science Foundation.} package.

In addition to the GMOS-N photometry and spectroscopy, we obtained near-infrared
(NIR) imaging of the two lensing clusters in the $\emph{z'JH}$ filters with the
Near-Infrared Camera and Fabry-Perot Spectrometer (NIC-FPS) of the 3.5m telescope
at the Apache Point Observatory (APO) in New Mexico. The detector
is a Rockwell Hawaii 1-RG 1024x1024 HgCdTe device with a pixel scale of $0.273\arcsec$
pixel$^{-1}$ and a $4.58\times4.58$ arcmin unvignetted field of view. The APO/NIC-FPS
$z'$ data differs from the Gemini/GMOS $z$ significantly due to the different
wavelength responses of the NIC-FPS and GMOS detectors. The effective wavelengths
of the two filters are offset by $\sim1000$\AA, and we use the prime ($'$)
throughout this paper to distinguish the APO/NIC-FPS $z'$ from the Gemini/GMOS $z$.
The NIR data were taken on 3 different nights in the Winter and Spring
of 2009. The conditions during the observing nights varied, with sub-arcsecond
seeing for the SDSS J1343+4155 $\emph{z'-}$band and both $\emph{J-}$band images.
The SDSS J0915+3826 $\emph{z'-}$band and both $\emph{H}$-band images were taken
in $\sim2\arcsec$ seeing. The observations consisted of 5-point dithers around a
$40\arcsec$ box and were reduced, registered and stacked using a custom IRAF pipeline.
Total exposure times are 5400s, 6960s and 3500s for $\emph{z'-}$, $\emph{J-}$, and
$\emph{H}$-band observations of SDSS J0915+3826 and 7800s, 5160s and 2625s for
$\emph{z'-}$, $\emph{J-}$, and $\emph{H}$-band observations of SDSS J1343+4155.

The SDSS was used to calibrate the four optical bands and the Two Micron All
Sky Survey (2MASS) was used to calibrate the NIR observations. Prior to making
any photometric measurements we first transform images in all bands to the
reference frame of the $\emph{i-}$band image, $0.1454 \arcsec$ pixel$^{-1}$ (GMOS-North
detector, binned 2x2). We then construct an empirical, normalized point spread
function (PSF) for each image based on a well-defined, non-saturated reference
star. We create photometric apertures by drawing a ridge line that covers the
high-redshift LAE and convolving it with the appropriate PSF for each image.
Apertures are defined by an equivalent radius, which corresponds to the radius
of a circular aperture that goes out to the same isophot. We make the final
magnitude measurement using a detailed sequence of sky subtraction and outlier
masking steps; first we subtract a general sky measurement and then compute the
median pixel value and standard deviation inside annuli of fixed width at increasing
radial distances from the source. Far enough out, these median values converge
to zero for an accurate sky subtraction. We average the median values
at large radii and subtract this average from the image to correct the general
sky subtraction. Outliers are defined as pixel values that deviate by more than
5$\sigma$ from the median in the respective annuli, and are replaced by the
median value plus an appropriate noise term. The final magnitude is measured at
an equivalent radius of twice the FWHM of the PSF and corrected to an equivalent
radius of $6\arcsec$ based on the curve of growth of the PSF. By defining the radius
as twice the FWHM, we make sure our apertures always cover the same physical region
on the sky. In the case of SGAS J091541+382655, a foreground galaxy lies very
close to and partially on top of the LAE galaxy. We use the $\tt{GALFIT}$ package
\citep{Peng:02} to fit a Sersic profile to this galaxy in the imaging data where
the LAE has no measureable flux, and then scale the galaxy model by the peak flux
in each band and subtract it out before measuring the LAE magnitudes.

When the LAEs are not detected (in $gJH$ for SGAS J091541+382655 and $grJH$ for
SGAS J134331+415455), we measure a limiting magnitude, derived from the total
sky noise in the aperture. We define a sky noise per pixel $\sigma_{sky}$ as the
standard deviation of our overall sky measurement plus a contribution from the
poison error made in this measurement. A pixel value of $2\sigma_{sky}$ is added
in quadrature for each pixel in our final aperture with an equivalent radius equal
to twice the FWHM. The limiting magnitudes are also aperture corrected to an
equivalent radius of $6\arcsec$. This method constrains the source to be fainter than the
limiting magnitudes at $95\%$ confidence in the relevant bandpasses.

All spectroscopic observations were carried out using the Gemini Multi-Object
Spectrograph \citep{GMOS} using custom slitmasks that were designed to target
lensed sources based on their color, morphology, and positoin in the GMOS imaging.
After targeting
all of the arc candidates, any remaining slits were placed on cluster members,
easily identified by their red sequence colors. Spectra were taken using
the macroscopic nod-and-shuffle (N\&S) mode available on GMOS. The reasons
for using N\&S are threefold. First, many of the emission or absorption
features that are used to determine galaxy redshifts in the range $z=1.0-3.0$
characteristic of the giant arcs, are in the redder part of the optical where
sky lines are problematic, and N\&S facilitates
more accurate sky-subtraction \citep{glaze2001}, particularly at low spectral
resolution.  Second, N\&S sky-subtraction allows us to use very small
$1\arcsec\times1\arcsec$ microslits that can be densely packed into the cluster core
($\sim 30\arcsec$), allowing us to target as many arcs, arclets, and cluster
galaxies as possible \citep{gilbank2008} . Third, as we are primarily interested in a
number of objects around the cluster center, the density of objects and the
limited field of interest are perfect for block-shuffling N\&S observations.
Spectra were taken with the R150\_G5306 grating in first order which gives a
dispersion of $3.5$\AA~ per pixel (binned spectrally by two), with six pixels
per resolution element resulting in a spectral FWHM~$\simeq 940$ km s$^{-1}$. Although
the R150 grating offers broad spectral range from the atmospheric cutoff to $\lambda
\gtrsim 1\mu{\rm m}$, the drop in sensitivity at the blue and red extremes, due
both to the GMOS CCD and the R150 grating efficiency, results in effective spectral
coverage of $\sim4000-9500$\AA~.

The N\&S technique employed in our program is nonstandard in that it involves a
nod distance on the sky that is half the size of the macroscopic shuffle. The
mask is designed to incorporate two submasks, each of which is a set of slits
covering an area one third the size of the detector. Slits for the two submasks
overlap on the sky in an area that is one sixth the size of the detector (because the
nod distance is set to half the shuffle distance). This design allows us to place
science slits for the primary target -- the core of a strong lensing cluster in
this case -- on both submasks and obtain useful spectra for the entire duration of
the N\&S exposure, whereas standard macroscopic N\&S results in science spectra for
only half of the total exposure time. Our integration times were 2400s resulting in
a 1200s effective integration for each of the two submasks. Two exposures were taken
for each target. Thus if an arc was targeted on both submasks (typical for the most
prominent arcs) the total integration time was 4800s. N\&S facilitates
straightforward sky subtraction by differencing two sections of the detector -- each
1/3 the size of the full detector. The spectra presented here
were wavelength calibrated, stacked, extracted and analyzed using a custom
pipeline based on the XIDL software
package\footnote{http://www.ucolick.org/$\sim$xavier/IDL/index.html}.

\subsection{SGAS J091541+382655}

SDSS J0915+3826 is a new strong lensing cluster that was discovered in the SGAS --
discussed above. We measure redshifts for 16 cluster members in our Gemini/GMOS
spectroscopy and combine this new data with a redshift for the BCG measured in
the SDSS to identify a mean cluster redshift of $z=0.397$ and a velocity
dispersion of $870\pm169$ km s$^{-1}$. The most prominent strong lensing feature
around this cluster is a bright blue arc that we identify as a background galaxy at
$z=1.501$ from O[II]$\lambda3727$\AA, CII]$\lambda2328$\AA,
CIII]$\lambda1907,1909$\AA~in emission, as well as AIII$\lambda1855,1863$\AA~in absorption.
Pre-imaging of this cluster reveals a source near the cluster core that exhibits
a dramatic drop in flux from the $\emph{i-}$ to $\emph{r-}$band, as shown in
Figure~\ref{dropout}. With Gemini/GMOS-N, in $\emph{griz}$,
we measure this source at $g_{AB}\geq26.07$, $r_{AB}=24.86\pm0.25$,
$i_{AB}=23.34\pm0.09$, and $z_{AB}=23.29\pm0.13$.
Slits were placed on this source in both submasks for our N\&S observation
of this cluster and the spectra exhibit bright emission at $7539$\AA~
which we interpret as Lyman-$\alpha$ $\lambda 1216$\AA~at $z=5.200$.

The LAE galaxy in this case is located on the sky very near to a foreground galaxy, and
so we took care to account for possible contamination in the LAE spectrum by light
from the galaxy that falls into the slit, as this could mimic a continuum break in the
source spectrum. The separation between the LAE box slit
and the potentially contaminating foreground galaxy is $2.5\arcsec$. Examination
of the radial profile of the potential contaminant in the pre-imaging observation
shows that the $i$-band flux within the LAE slit due to the galaxy is on average
1.5$\times$10$^{-3}$ of the peak flux of that galaxy. The mean $i$-band flux of the
LAE within the spectroscopic aperture is 5.1$\times$10$^{-2}$ compared to the peak
of the contaminating galaxy, and thus assuming comparable seeing between the imaging
and spectroscopic observations suggests that this object's contribution to the $i$-band
flux within the LAE aperture is approximately $3\%$ of the total light. Note that
given the separation between the contaminant and the slit the seeing of the
spectroscopic and imaging observations would have to be grossly different to
produce significant contamination; such a mismatch is neither expected due to
execution of the observations in specified conditions in queue mode, nor suggested
by the spatial width of stellar spectra within the alignment boxes of the
spectroscopic observations.

\begin{figure}
\centering
\includegraphics[scale=0.51]{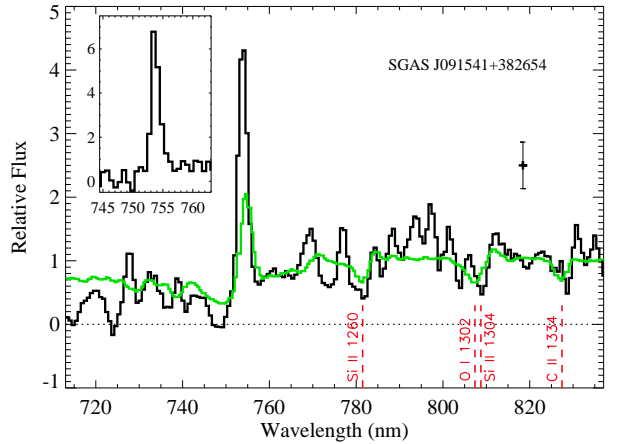}
\caption{\scriptsize{GMOS spectrum of the Lyman-$\alpha$ emitter behind
cluster lens SDSS J0915+3826. The LAE spectrum plotted over a range of
$\sim7150-8350$\AA~is smoothed to a scale comparable to a the spectral
resolution of $940$ kms$^{-1}$ in order to emphasize the low-significance
candidate absorption features, with the composite Lyman Break Galaxy
absorption+emission spectrum from \citet{shapley2003} overplotted in
green. Dashed vertical lines indicate the location of prominent UV
continuum metal absorption lines at the LAE redshift, which
coincide with the low-significance features in our continuum spectrum.
The Lyman-$\alpha$ emission feature is also shown inset, without smoothing.
A typical spectral flux error bar is displayed on the right side of the figure,
in the whitespace above the continuum.}}
\label{spec0915}
\end{figure}

The relevant part of the extracted spectrum is displayed in Figure \ref{spec0915},
and despite the low dispersion grating used for our observations the line is measurably
asymmetric. We also detect continuum signal redward of the emission line and no
significant signal blueward of it. This source has a strong photometric $\emph{r-i}$
break -- even after subtracting the emission line flux from the $\emph{i-}$band
photometry. It is also very blue in $\emph{i-z}$, and is not detected in NIC-FIPS
imaging to a 2-$\sigma$ limiting magnitude of $J_{AB}\geq22.49$, indicatingg that
the continuum spectral slope redward of the emission feature is blue. The blue
continuum redward of the emission and the strong $\emph{r-i}$ break --
along with the non-detection in $\emph{g}$, and the asymmetry of the line confirm the
interpretation of the source as a Lyman-$\alpha$ emitting galaxy at high redshift. The
most likely candidate emission lines which confuse searches for high-redshift LAEs
are O[II]$\lambda 3727$\AA, [OIII] $\lambda 5007$\AA~and H-$\alpha$, but for each of
these other candidate lines we would expect accompanying emission features to fall
within the wavelength coverage of our spectroscopy. We measure the equivalent width
of the Lyman-$\alpha$ emission by fitting a simple
power-law to the continuum spectrum redward of the emission line over a wavelength range
of $7575-8275$\AA, and subtracting the underlying continuum flux from the emission line.
Over the relatively small wavelength range used the continuum level is consistent with
a constant value. We note that our spectral flux calibration uses a standard from the
Gemini archive to correct roughly for the detector sensitivity as a function of
wavelength, but is not suitable for measuring absolute fluxes from the spectral
data. Taking into account
the underlying continuum and the source redshift we measure a rest-frame EW for the
Lyman-$\alpha$ emission of $25\pm4$\AA, and combine the EW measurement with our
$i-$band photometry to calculate an observed Lyman-$\alpha$ line flux of
$2.1\pm0.7\times$10$^{-16}$ ergs s$^{-1}$ cm$^{-2}$.

\begin{figure}
\centering
\includegraphics[scale=0.51]{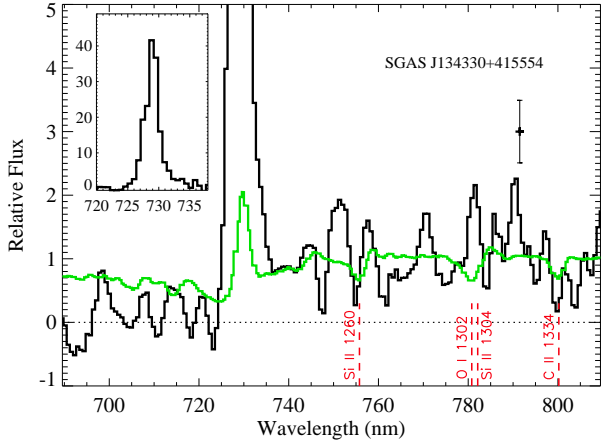}
\caption{\scriptsize{GMOS spectrum of the Lyman-$\alpha$ emitter behind
cluster lens SDSSJ1343+4155. The spectrum is presented identically to
the spectrum shown in Figure~\ref{spec0915}, except that it covers an
observer-frame wavelength range of $\sim6900-8100$\AA.}}
\label{spec1343}
\end{figure}

\subsection{SGAS J134331+415455}

The second lensed LAE presented here is located near the strong lensing cluster
SDSS J1343+4155, which was previously identified as a strong lensing cluster in
the SDSS by \citet{wen2009} and \citet{diehl2009}. We combine our Gemini/GMOS
spectroscopy with two cluster member redshifts from the SDSS to measure a mean cluster
redshift of $z=0.418$ and a velocity dispersion of $1077\pm266$ km s$^{-1}$ from 7
cluster member galaxies. The bright arc around this lens is spectroscopically
identified by \citet{diehl2009} as a background galaxy at z$=2.091$. Pre-imaging
of this cluster reveals a source near the cluster core that exhibits a dramatic drop
in flux from the $\emph{i-}$ to $\emph{r-}$band -- shown in Figure~\ref{dropout}. We
measure this source at $g_{AB}\geq25.97$, $r_{AB}\geq25.64$,
$i_{AB}=23.78\pm0.12$, and $z_{AB}=24.24^{+0.18}_{-0.16}$.  We placed slits
on this source in each of the two submasks for our spectroscopic N\&S observations of
this cluster. The spectra corresponding to these slits exhibit a bright emission line at
$7289$\AA, which we interpret as Lyman-$\alpha$ $\lambda 1216$\AA~at $z=4.994$.

The extracted spectrum around the emission feature is displayed in
Figure~\ref{spec1343}. The emission feature is significantly asymmetric and we
measure continuum emission redward of the line, but no continuum blueward.
The source is undetected in NIC-FIPS imaging down to a 2-$\sigma$ limiting magnitude
of $J_{AB}\geq22.22$. Similarly to the case of SGAS J091541+382655, this source
has a blue $\emph{i-z}$ color and our NIR photometry implies that it has a blue continuum
spectral slope redward of the emission feature. The most likely candidate emission
line which could be misinterpreted as Lyman-$\alpha$ in this spectrum is O[II]
$\lambda 3727$\AA, but we rule this out as a realistic interpretation based on the
absence of H-$\beta$ $\lambda 4862$\AA~and O[III] $\lambda 4960,5007$\AA, both of
which should fall just within the spectral range covered by our data if the source
were a very red galaxy at lower redshift. Other common contaminants in searches for
high-redshift LAEs include [OIII] $\lambda 5007$\AA~and H-$\alpha$, but in each of
these cases we would also expect to observe other accompanying emission features given
the wavelength coverage of our spectroscopy. We measure
the equivalent width of the Lyman-$\alpha$ emission by fitting a simple power-law
to the continuum spectrum redward of the emission line over a wavelength range
of $7400-8275$\AA, and subtracting the underlying continuum flux from the integrated
flux from the emission line. Similar to the continuum fit for SGAS J091541+382655,
the continuum level for SGAS J134331+415455 is consistent with a constant value.
Taking into account the continuum and the source redshift we measure a rest-frame EW for the
Lyman-$\alpha$ emission of $136\pm20$\AA~and an combine the EW measurement with our
$i-$band photometry to calculate an observed Lyman-$\alpha$ line flux of
$2.1\pm0.5\times10^{-16}$ ergs s$^{-1}$ cm$^{-2}$.

\section{Analysis}

\subsection{Lens Models and Intrinsic Source Properties}

\begin{figure}
\centering
\includegraphics[scale=0.3]{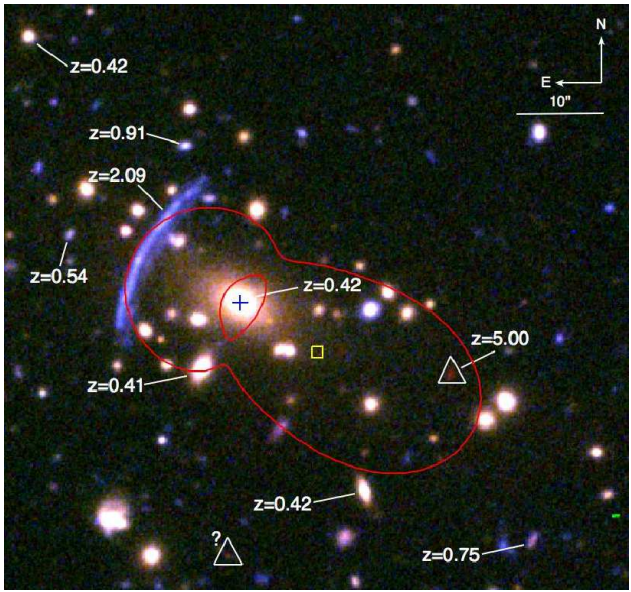}
\caption{\scriptsize{The field of SDSS J1343+4155 with tangential and
radial critical curves for the bright main arc overplotted. One image
of the lensed background LAE, spectroscopically confirmed at $z=4.994$,
and its putative counterimage predicted by the lensing model are
identified by white triangles. The center of the dark matter halo and
BCG mass components are indicated by a square and cross, respectively.
Other sources in the field -- background
objects and cluster members -- are labeled with their spectroscopic
redshifts. Color images are created from Gemini+GMOS-North $gri$ 300s
pre-imaging data.}}
\label{fig:mag1343}
\end{figure}

Using the measured equivalent widths and $\emph{i}$ magnitudes we also
calculate lensed isotropic Lyman-$\alpha$ line luminosities of
L$_{Ly-\alpha}=5.9\pm1.1\times10^{43}$ erg s$^{-1} h_{0.7}^{2}$ and
L$_{Ly-\alpha}=5.4\pm0.9\times10^{43}$ erg s$^{-1} h_{0.7}^{2}$
for SGAS J091541+382655 and SGAS J134331+415455, respectively. These are
not the true isotropic luminosities because both sources are lensed by foreground
galaxy clusters, and are therefore significantly magnified. To measure the
intrinsic luminosities of the LAEs, we estimate the magnification
due to strong lensing by the intervening clusters. The mass models are constructed
 using the publically available software, $\tt{LENSTOOL}$ \citep{jullo2007}, with
Monte Carolo Markov Chain (MCMC) minimization in the source plane, and are shown
in Figure~\ref{fig:mag1343}. For
SDSS J1343+4155, we compute a simple mass model using as constraints the
giant blue arc at $z=2.09$, the LAE at $z=4.994$, paired with a counter image
candidate that we have identified. The cluster and the brightest cluster galaxy
(BCG) are represented by pseudo isothermal elliptical mass distributions
\citep[PIEMD;][]{jullo2007}. We allow all the parameters of the cluster
halo to vary. For the BCG, we follow the light distribution for the
positional parameters, and vary the core and cut radii and the
velocity dispersion. Our best-fit model reproduces the locations and
orientations of the observed lensed images, and is consistent with the
measured velocity dispersion to within the uncertainty. Based on this model,
the magnification at the location of the LAE is
$m\sim12$, which we use to calculate the intrinsic isotropic Lyman-$\alpha$
line luminosity for SGAS J134331+415455 to be
L$_{Ly-\alpha}=4.5\pm0.7\times10^{42}$ erg s$^{-1} h_{0.7}^{2}$.
Assuming Case B recombination and taking the prescription from \citet{kenn1998}
we calculate the star-formation rate (SFR) for SGAS J134331+415455 to be
SFR$_{Ly-\alpha}=4.1\pm0.6$ M$_{\sun}$ yr$^{-1} h_{0.7}^{-1}$. Applying a
mangification correction to the apparent magnitude in $z-$band we find
the intrinsic $z-$band AB magnitude to be $z=27$. We can also estimate
the SFR from the UV continuum flux density at $1500$\AA~according to equation
$(1)$ from \citet{kenn1998}, which results in SFR$_{UV}=18\pm3$ M$_{\sun}$ yr$^{-1} 
h_{0.7}^{-1}$.

\begin{figure}
\centering
\includegraphics[scale=0.3]{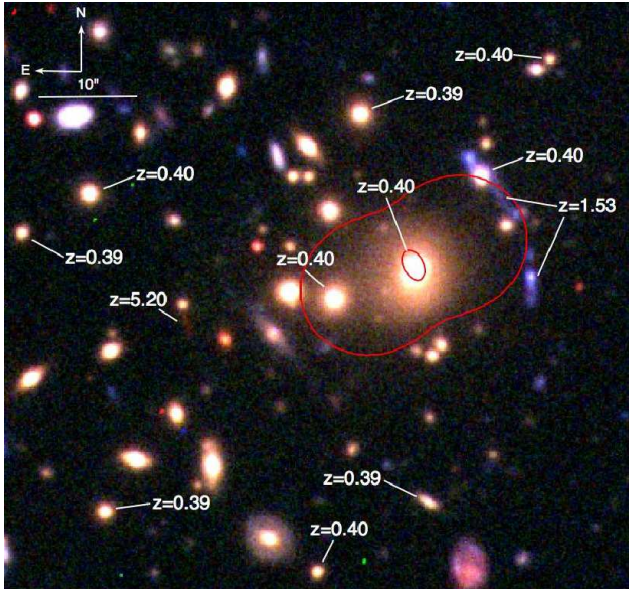}
\caption{\scriptsize{The field of SDSS J0915+3826 with the tangential
critical curve for the bright main arc overplotted. Other sources with
spectroscopic redshifts are also identified, including the lensed
background LAE at $z=5.200$. Color images are created from
Gemini+GMOS-North $gri$ 300s pre-imaging data.}}
\label{fig:mag0915}
\end{figure}

The strong lensing model for SDSS J0915+3826 is constrained by the positions
of the components of the giant arc. We do not identify a clear counter
image for the LAE, and therefore do not use it as a constraint. We represent the
lens with a single PIEMD and allow all its
parameters to vary, which produces a velocity dispersion estimate that
is similar to the value we measure from spectroscopy of cluster members
(see Section 2.2). The critical curves for our
best-fit model are plotted in Figure~\ref{fig:mag0915}.
Since the available constraints are internal to the radial projection
of the LAE, the predicted LAE magnification is highly uncertain.
In particular, if we compute a set of models with parameters in the
range allowed by the uncertainties, the location of the critical curve
for the LAE varies significantly and there is practically no upper limit for the
implied magnification. Since in all cases the magnification is higher
than $\sim10$, we adopt this number as our working order-of-magnitude estimate
and determine an upper limit for the intrinsic isotropic Lyman-$\alpha$
line luminosity for SGAS J091541+382655 of L$_{Ly-\alpha}\leq5.9\pm1.1\times10^{42}$
ergs s$^{-1} h_{0.7}^{2}$ and a corresponding SFR$_{Ly-\alpha}\leq5.3\pm1.0$
M$_{\sun}$ yr$^{-1} h_{0.7}^{-1}$. Correcting the $z-$band magnitude for the lensing
mangifciation we recover an intrinsic magnitude of $z\geq25.75$ mags, from which
we measure SFR$_{UV}\leq51\pm7$ M$_{\sun}$ yr$^{-1} h_{0.7}^{-1}$ using equation
(1) from \citet{kenn1998}.

We note that SFR$_{Ly-\alpha}$ and SFR$_{UV}$ differ significantly for both sources.
For SGAS J134331+415455 SFR$_{Ly-\alpha}$ is lower than SFR$_{UV}$ by a
factor of $4.5$, which is consistent with previous studies finding that the
SFR calculated from Lyman-$\alpha$ emission is often systematically lower
than other SFR metrics -- such as UV continuum flux density at 1500\AA~--
by a factor of $\sim5$ or more \citep{tapken2007}. The disagreement between
the two SFR estimates is more extreme for SGAS J091541+382655, with SFR$_{UV}$
$\sim9\times$ larger than SFR$_{Ly-\alpha}$. In a study of LAEs at $z=3.1$ identified
in the Chandra Deep Field South, \citet{gronwall2007} find on average,
SFR$_{UV}$/SFR$_{Ly-\alpha}\sim3$ with a large scatter about the
mean, including several objects with SFR$_{UV}$/SFR$_{Ly-\alpha}$ as large as we
measure for SGAS J091541+382655. Significant differences between these two SFR
calibrations are understandable considering the potential systematic errors
involved in measuring the SFR from both UV continuum and Lyman-$\alpha$ emission.
For example, resonant scattering of Lyman-$\alpha$ photons off of
neutral hydrogen can supress observed Lyman-$\alpha$ emission if there is even
a very small amount of dust in the source galaxy, thus producing an underestimate of
SFR$_{Ly-\alpha}$. Additionally, \citet{kenn1998} identifies several caveats associated
with the SFR$_{UV}$ calibration, including assumptions about the IMF and the timescale
and manner of star-formation (e.g. constant vs. burst). Specifically, equation $1$ in
\citet{kenn1998} recovers SFR from the rest-frame UV flux by assuming a galaxy
has undergone continuous star formation for $10^{8}$ years or longer, and originates
from a Salpeter IMF with mass limits of $0.1$ and $10$ M$_{\sun}$. If the properties
of the underlying stellar populations in either of these galaxies differ significantly
from the parameters assumed in Equ. $1$ from \citet{kenn1998} then there will be
systematic differences between the SRFs estimated from the rest-frame UV flux and
the true SFRs.

We have also considered the possibility that the Lyman-$\alpha$ emission in one or both
of these objects could be due all or in part to AGN activity. Given the lack of
detection of significant emission from highly ionized species, such as N[V] and
C[IV], we conclude that it is unlikely that AGN are playing a significant role
in the observed Lyman-$\alpha$ emission. The disagreements between
the two SFR estimates for the two sources discussed here highlight the difficulty
in making robust SFR measurements from source-frame UV observables.

Having accounted for the magnification due to gravitational lensing, we can
also compare the luminosities of these two sources to large
samples of LAEs and LBGs at comparable redshifts
\citep[e.g.,]{hu2004,shimasaku2006,bouwens2007,dawson2007,murayama2007,ouchi2008,mclure2009}.
Studies of luminosity functions at high redshift are challenging,
and parameters such as L$^{*}$ and $\alpha$ are not nearly so well measured as they
are in the nearby universe. For comparing Lyman-$\alpha$ luminosities, we take the average
of 6 different measurements of L$_{Ly-\alpha}^{*}$ values from Ouchi et al. (2008)
that were fit using samples of LAEs at $z\sim3.7$ and $z\sim5.7$, assuming three
possible values of the faint end slope of the luminosity function, $\alpha$. For the
purpose of comparing continuum luminosities, we take L$_{UV}^{*}$ as the average of
the two L$_{1500}^{*}$ best-fit values for LAEs at $z\sim3.7$ and $z\sim5.7$ from
\citet{ouchi2008}. SGAS J091541+382655 is
$($L$_{Ly-\alpha}^{*}$, L$_{UV}^{*}) \leq (0.6$L$_{Ly-\alpha}^{*}$, $2$L$_{UV}^{*})$,
and SGAS J134331+415455 is
$($L$_{Ly-\alpha}^{*}$, L$_{UV}^{*}) = (0.5$L$_{Ly-\alpha}^{*}$, $0.9$L$_{UV}^{*})$.
Both of these galaxies live at $\lesssim$L$^{*}$ on the luminosity function for
similar galaxies at comparable redshifts, which makes them interesting targets for
studying the properties of typical LAEs at $z\sim5$ on an individual basis.

We also attempt to measure the morphologies of both lensed LAEs in our Gemini/GMOS
imaging and place constraints on the intrinsic sizes of these two galaxies. Neither
source is detected at $S/N\gtrsim12$, which limits our ability to make detailed
morophological measurements or shape fits, but we can check the object
profiles against the image PSFs and use our estimates of the magnification due to lensing to
place some rough constraints on the physical sizes of these galaxies. Studies of
LAE morphologies at $z=3.1$ using $HST$ imaging find that the Lyman-$\alpha$ and
restframe UV emission from these sources are spacially coincident and have half-light
radii $\lesssim 1.5$kpc \citep{bond2009,bond2010}. \citet{bond2009} argue that
S/N$\gtrsim30$ is necessary to make robust measurements of half-light radii for
LAEs, and find that it is difficult to distinguish between resolved and unresolved
compact cores at S/N$\lesssim30$. The two lensed LAEs are best-detected in our
Gemini/GMOS $i-$band imaging, so this is the data we use for our morphological
measurements.

SGAS J134331+315455 is slightly elongated in the tangential
direction with respect to the center of the cluster. We
measure its profile to be consistent with a gaussian along the tangential axis with
a FWHM$=1.1\arcsec$, and consistent with a gaussian of FWHM$=0.57\arcsec$ along the
radial axis. The FWHM of the PSF measured from stars in the
image is $\sim0.57\arcsec$, suggesting that the LAE is unresolved along one axis and
resolved along the other. This kind of morphology is natural for a strongly lensed
background source where the magnification is generally much greater along one axis
that the other; a source located on the tangential caustic will be highly magnified
in the tangential direction with respect to the center of the lensing potential, and
magnified very little or not at all in the radial direction. We conclude that
SGAS J134331+415455 is unresolved in the radial direction and barely resolved in the
tangential direction. Assuming that all of the magnification factor of $m=12$ is
applied in the tangential direction, we convert a FWHM of $1.1\arcsec$ into a
physical linear size of $\sim0.6$kpc, which indicates that most of the emission
from this galaxy originates from a region very compact in size. This constraint
must be taken with the caveat that the data are lower S/N than would be optimal,
but the signficant linear magnification of these sources due to gravitational
lensing does enable us to probe very physically interesting scales even with
ground-based imaging. We also measure the locations of SGAS J134331+415455 to
be identical in the Gemini/GMOS $i-$ and $z-$band images to the accuracy of our
astrometry, $\sim0.2"$, corresponding to physical scales of $\sim100 pc$. Based on
the observer frame equivalent width of this source, the flux in the
$i-$band filter is $\sim60\%$ Lyman-$\alpha$ line emission, whereas the
$z-$band filter measures only rest-frame UV continuum emission. Given the spatial
coincidence of the LAE in the $i-$ and $z-$bands we conclude that the UV continuum and
Lyman-$\alpha$ emission cannot originate from regions separated by more than
of order $\sim100 pc$ projected onto the sky. We have no measurement of the
relative velocities of the UV continuum and the Lyman-$\alpha$ emission and
therefore cannot constrain a possible separation in velocity.

SGAS J091541+382655 is similar to SGAS J134331+415455 in that
the LAE is elongated on the tangential axis relative to the center of the lensing
cluster. Using the $\tt{GALFIT}$ subtracted $i-$band image, we measure the LAE to be
resolved in the tangential direction, having a shape consistent with a gaussian
of FWHM$=1.82"$, which equates to a constraint on the physical scale of $\leq1.2$kpc,
where the lower limit on the magnification of this source translates to an upper limit
in the physical size (and again assuming that the magnification is entirely along the
tangential diretion). We also measure the LAE to have a gaussian shape with
$FWHM=0.75\arcsec$ in the radial direction, which -- given the FWHM$\sim0.73\arcsec$
measured from the PSFs of stars in the image -- is consistent with the source being unresolved
in the radial direction. In the observer frame the equivalent width of the
Lyman-$\alpha$ emission for this source is EW$_{Ly-\alpha}=157$\AA, from which
we estimate that Lyman-$\alpha$ photons are contributing less than $20\%$ of
the $i-$band flux, so that we cannot make any attempt to measure a spatial offset
between the Lyman-$\alpha$ and
UV continuum emission. Our size constraints for both sources are in good agreement
with the literature, in that we find the sizes of the LAEs to be consistent with
compact sources of size $\lesssim 1.5$kpc.

\begin{figure}
\centering
\includegraphics[scale=0.51]{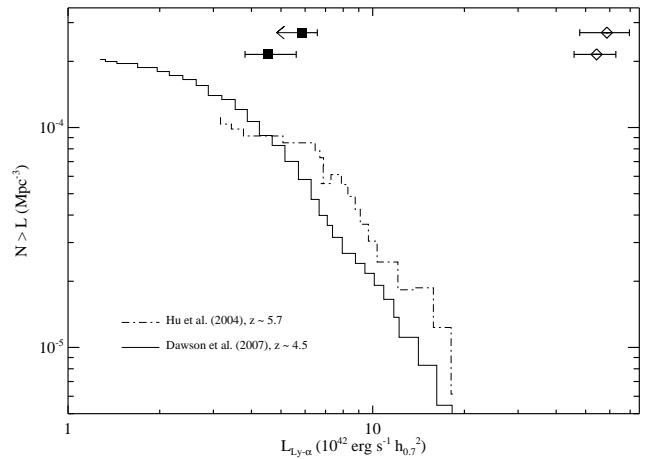}
\caption{\scriptsize{Two empirical cumulative Lyman-$\alpha$ line emission
luminosity functions (uncorrected for completeness) for LAEs at high redshift
are plotted; a sample of $z\sim4.5$ LAEs published by Dawson et al.(2007) is
given by the solid histogram and a sample of $z\sim5.7$ LAEs \citep{hu2004}
by the dot-dashed histogram. Both the apparent (magnified) and intrinsic
Lyman-$\alpha$ luminosities for the
two sources presented in this paper. Open diamonds indicate the apparent luminosities
and filled squares indicate the intrinsic luminosities, corrected for the magnification
(where we have only an upper limit to the intrinsic luminosity for SGAS
091541+382655). The magnification due to strong lensing
allows us to probe deep down the Lyman-$\alpha$ luminosity function
with sources that are observationally the brightest two LAEs known at
$z\sim5$.}}
\label{laelf}
\end{figure}

\begin{figure}
\centering
\includegraphics[scale=0.51]{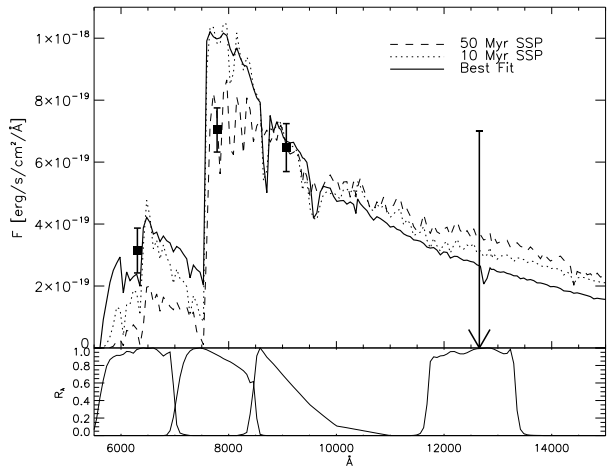}
\caption{\scriptsize{The best-fit spectral energy distribution for
SGAS J091541+382655 is plotted with a solid line on top of the
photometric data. The dotted and dashed lines are
dust-free, single burst models from CB07, with ages of 10\,Myr
and 50\,Myr respectively, scaled to match the observed
$z$-band flux. This illustrates that the photometry favors a
young (i.e. blue) stellar population, the 50\,Myr model
already deviates considerably from the best-fit SED. Filter
transmission curves corresponding to the $r$, $i$, $z$ and $J$
photometry are plotted in the bottom panel.}}
\label{fig:sed0915}
\end{figure}

\begin{figure}
\centering
\includegraphics[scale=0.51]{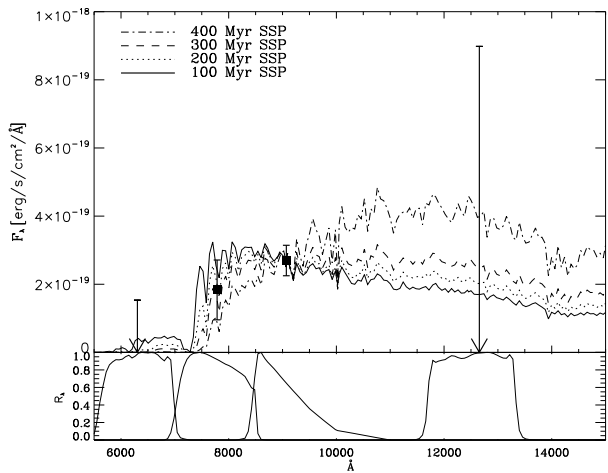}
\caption{\scriptsize{Dust-free single burst CB07 models are scaled to
match the $z$-band flux and plotted on top of the photometric
data for SGAS J134331+415455. Models with ages anywhere in the
range of 100\,Myr to 400\,Myr are potentially good fits for the
limited data in-hand. Filter transmission curves corresponding
to the $r$, $i$, $z$ and $J$ photometry are plotted in the
bottom panel.}}
\label{fig:sed1343}
\end{figure}

\subsection{Stellar Mass and UV Continuum Properties}

With several photmetric measurements in hand for each of our
sources it becomes interesting to investigate the spectral
energy distributions (SEDs) of the LAEs to try and recover the
properties (e.g. stellar mass, age, SFR, and dust content) of
the underlying stellar populations in a way that does not rely
on as many problematic assumptions as, for example, the SFR estimated
from the rest-frame UV continuum flux that we calculate in Section 3.1
above. We compare our photometry for these sources against modified stellar
population synthesis models. More details can be
found in \citet{wuyts2010}, we only summarize the main
procedure here. We use the revised templates of Bruzual and Charlot
(CB07, based on \citet{bruzchar03}) with solar metallicity, a
Chabrier initial mass function \citep{chabrier2003} and
\citet{calzetti2000} dust extinction
law. We investigate a range of exponentially declining star
formation histories of the form $SFR(t) \sim \exp(−t/\tau)$,
with $\textit{e}-$folding times $\tau = $0.01, 0.05, 0.1, 0.2,
0.5, 1, 2 and 5 Gyr, as well as single bursts (SSP) and
continuous star formation models (CSF). An updated version of
the code $\textit{Hyperz}$ \citep{Bolzonella:00} is used to
obtain the best-fit SED at the
fixed spectroscopic redshift of the source via a maximum
likelihood procedure.

The best-fit SED for SGAS J091541+382655 is shown in Figure
\ref{fig:sed0915} and corresponds to a dust-free single burst with an age
of 1.4\,Myr. It is important to supplement the best-fit
stellar population parameters with confidence intervals
allowed by the photometric uncertainties. We create 1000 fake
realizations of the observed SED by perturbing each
broadband magnitude measurement in a manner consistent with
its errorbars. This set of fake SEDs is fit in exactly the
same manner as described above for the observed SED; bad fits
with $\chi^{2} > 3$ are excluded. This procedure results in a
very young (age $\leq$ 5\,Myr) and dust-free ($E(B-V) \leq 0.01 $)
stellar population. Though we can place constraints on the age of
the observed stellar population, are data are not sufficient for the
SED fits to simultaneously constrain the star formation history, and
consequently the current star formation rate, for this galaxy. After
correcting for a lensing magnification factor of $\gtrsim10$ , we find
a Chabrier stellar mass of M$_{stars} \leq 7.9^{+3.7}_{-2.5} 
\times 10^{7}$ M$_{\sun} h_{0.7}^{-1}$.

SGAS J134331+415455 is detected in only 2 bands, which is
insufficient to produce a robust best-fit stellar population.
Instead, we compare the photometry to single burst, dust-free
CB07 models of different ages to obtain some constraints on
the age and stellar mass. The results are shown in Figure
\ref{fig:sed1343}. The age of the observed stellar population lies in the range
between 100\,Myr and 400\,Myr, which makes SGAS J134331+415455
significantly older than the very young stellar population
seen in SGAS J091541+382655. After correcting for a lensing
magnification factor of 12, we constrain the corresponding stellar masses
to be within the range $2\times10^{8}$ M$_{\sun} h_{0.7}^{-1} <$
M$_{stars} < 6\times10^{9}$ M$_{\sun} h_{0.7}^{-1}$. All of our
photometric data for both of these sources sample light
blueward of the 4000\AA~break in the source frame so that the
stellar population parameters obtained from the SED fitting
procedures correspond to the most recent episode of star formation
in each LAE; we have no power to constrain stellar mass that may be present
in an underlying population of older stars.

In addition to SED modeling of the photometry of both LAEs, we also examine the modest
continuum signal that is detected for SGAS J091541+382655. Cross correlation of our
GMOS spectrum for this source over a wavelength range of 1230-1350\AA~(source frame)
against the composite LBG spectrum from \citet{shapley2003} results in a
peak at $z\sim5.2$, in agreement with the redshift measured from the Lyman-$\alpha$
emission. We explicitly exclude the Lyman-$\alpha$ emission portion of the spectrum
in this analysis so that the cross-correlation signal that we detect is entirely due
to low-significance continuum features redward of Lyman-$\alpha$. The composite LBG
from \citet{shapley2003} is plotted with our GMOS spectra for both targets in
Figures~\ref{spec0915} and \ref{spec1343}, and visual comparison of the continuum
against the LBG composite spectrum suggests the
presence of several strong UV metal absorption lines that are observed in well-studied LBGs
at lower redshifts \citep[e.g.][]{pettini2000}. The features are at too low a significance
in our data to claim a robust detection of individual absorption lines, but the cross
correlation signal alone is encouraging given our low resolution
spectra and limited integration time. Both of the sources presented here
are excellent candidates for a more aggressive spectroscopic followup effort to
explore the gas properties and stellar metallicity in representative low-mass starforming
galaxies at $z\sim5$.

\section{Summary and Conclusions}

We have identified two lensed Lyman-$\alpha$ emitting galaxies at $z\sim5$ near
the cores of strong lensing selected galaxy clusters. These sources are
among the brightest galaxies identified at such high redshift, but their intrinsic
luminosities are much lower than the observed flux due to magnification by
the gravitational potential of foreground galaxy clusters. We use the available
data to investigate the underlying stellar populations for these galaxies and
find that the light -- continuum and line emission -- for SGAS 091541+382655
likely originates from a population of young stars with low dust content.  Both sources
are in the process of undergoing active star formation. Our analysis of these two
LAEs corroborates our current understanding of the nature of Lyman-$\alpha$ emitting
galaxies at high redshift,
and the large magnification of these sources due to gravitational lensing makes
them excellent candidates for studying the individual properties of galaxies
on the faint end of the L$_{Ly-\alpha}$ and L$_{UV}$ luminosity functions at $z\sim5$.
We encourage efforts to followup these sources
aggressively on 8-10m class telescopes in order to better study the properties of the
underlying stellar populations via continuum light. These sources are also excellent
targets for space-based observations, both with current $HST$ instruments and with
$JWST$ in the future.

\acknowledgments{We thank the anonymous referee for exceptionally thoughtful
and helpful comments which helped to significantly improve the quality of this
paper. MBB acknowledges the support of the Sigma Xi Scientific Research Society
in the form of a grant in-aid of research. We also wish
to thank the Gemini North observing and support staff for their efforts
in taking data that contributed to this paper. The authors wish to recognize
and acknowledge the very significant cultural role and reverence that the
summit of Mauna Kea has always had within the indigenous Hawaiian community.
We are most fortunate to have the opportunity to conduct observations from
this mountain.}

\bibliographystyle{apj}

\clearpage

\end{document}